\documentstyle[12pt]{article}
 \advance\oddsidemargin by -0.9in 
\advance\evensidemargin
 by -0.9in 
\advance\topmargin by -0.10in 
\makeindex

\def\lsim{\mathrel{\rlap{\lower4pt\hbox{\hskip1pt$\sim$}}
    \raise1pt\hbox{$<$}}}         
\def\gsim{\mathrel{\rlap{\lower4pt\hbox{\hskip1pt$\sim$}}
    \raise1pt\hbox{$>$}}}         
\def\beq{\begin{equation}}
\def\endeq{\end{equation}}
\def\arr{\begin{eqnarray}}
\def\endarr{\end{eqnarray}}
\def\Pom{{\bf I\!P}}

\begin{document}
\vspace*{0.15cm}
\hspace*{9.65cm} {\Large \bf DFTT 20/96} \\
\hspace*{10.3cm} {\Large \bf hep-ph/9605208} \\
\vspace*{1.50cm} 

\centerline{\huge \bf
Anomalies in diffractive }
\medskip
\centerline{\huge \bf
electroproduction of 2S radially}
\smallskip
\centerline{\huge \bf 
excited light vector mesons}
\smallskip 
\centerline{\huge {\bf
at HERA}\footnote{Talk presented by J.
Nemchik at the International 
Conference HADRON STRUCTURE '96, \\ Star\' a Lesn\' a,
February 12-16, 1996, Slovak Republik}}

\vspace{0.60cm}

\begin{center}
{\large \bf
{J.~Nemchik}} 

\vspace{0.1cm}

{\sl Dipartimento di Fisica Teorica, Universit\` a di Torino,\\
and INFN, Sezione di Torino, I-10125, Torino, Italy}

\vspace{0.2cm}

{\sl Institute of Experimental Physics, Slovak Academy of Sciences,\\
Watsonova 47, 04353 Kosice, Slovak Republik} 

\vspace{0.3cm}

{\large \bf N.N.~Nikolaev}

\vspace{0.1cm}

{\sl IKP(Theorie), KFA J{\"u}lich, 5170 J{\"u}lich, Germany}

\vspace{0.2cm}

{\sl L. D. Landau Institute for Theoretical Physics, GSP-1,
117940, \\
ul. Kosygina 2, Moscow 117334, Russia.} 

\vspace{0.3cm}

{\large \bf  E.~Predazzi} 

\vspace{0.1cm}

{\sl Dipartimento di Fisica Teorica, Universit\` a 
di Torino,\\
and INFN, Sezione di Torino, I-10125, Torino, Italy.} 

\vspace{0.3cm}

{\large \bf B.G.~Zakharov} 

\vspace{0.1cm}

{\sl L. D. Landau Institute for Theoretical Physics, GSP-1,
117940, \\
ul. Kosygina 2, Moscow 117334, Russia.} 
\end{center}

\vspace{0.3cm}

\begin{abstract}

We present the color dipole phenomenology
of diffractive photo-
and electroproduction $\gamma^{*}\,N\rightarrow V(V')\,N$ of
vector mesons ($V(1S) = \phi^0, \omega^0, \rho^0, J/\Psi, \Upsilon$) 
and their radial
excitations ($V'(2S) = \phi', ~\omega', ~\rho', 
~\Psi', ~\Upsilon'$). 
The main emphasis is related to light vector mesons.
We discuss how 
the energy dependence of the color dipole cross section
in conjuction with the node
of the radial wave function of the $2S$ states 
can lead to an anomalous $Q^2$ and energy dependence of
diffractive production of $V'(2S)$ vector mesons.
The color dipole model predictions 
for $V(1S)$ light vector meson production are compared
with the experimental data from the 
EMC, NMC, ZEUS and H1 collaborations. 

\end{abstract}

\newpage
\setlength{\baselineskip}{0.76cm}
 
 
\section{What can we learn from diffractive
         electroproduction of vector mesons ?}
 
~~~~One of the important feature of
diffractive electroproduction of vector mesons
\beq
\gamma^{*}p\rightarrow Vp\,,~~~~~V=\rho^{0},\,\omega^{0},\,
\phi^{0},\,J/\Psi,\,\Upsilon\,... 
\label{eq:1.1}
\endeq
at high energy, $\nu$,
is a possibility to study 
the pomeron exchange
\cite{DL,KZ91,Ryskin,KNNZ93,KNNZ94,NNZscan,Brodsky,Forshaw}.
The high energy hadrons and photons are considered as color
dipoles in the mixed $({\bf{r}},z)$
lightcone representation \cite{NZ91,NZ94}
with the transverse size, ${\bf{r}}$, frozen during the interaction process.
The interaction (scattering) process
is characterized 
by the color dipole cross section, $\sigma(\nu,r)$, which
represents the
interaction of color dipoles with the target nucleon.
The energy evolution of the color dipole cross section 
is described by the generalized BFKL (gBFKL)
equation \cite{NZ94,NZZ94}.
The
$V(1S)$ vector meson production amplitude probes the color dipole cross
section at the dipole size, $r\sim r_{S}$, where $r_{S}$ is the scanning
radius. This property is the so-called
{\it scanning phenomenon} \cite{NNN92,KNNZ93,KNNZ94,NNZscan} and
reflects the 
shrinkage of the transverse size of the virtual photon with
$Q^{2}$ together with the small-size
behaviour of the dipole cross section ($\sim r^{2}$).
The scanning radius can be expressed through the scale
parameter, $A$, as
\beq
r_{S} \approx {A \over \sqrt{m_{V}^{2}+Q^{2}}}\, .
\label{eq:1.2}
\endeq
At large $Q^{2}$ and/or for heavy vector mesons,
the scanning radius is small and
the production amplitude of reaction (\ref{eq:1.1}) is 
perturbatively calculable. 
However, due to a large scale parameter, 
$A\approx 6$, in (\ref{eq:1.2}) \cite{NNZscan},
the onset of the short-distance dominance is very slow
for the production of light vector mesons even
at the moderate $Q^{2} \lsim 20$\,GeV$^{2}$
corresponding to the present fixed target and HERA experiments.
Therefore, changing $Q^{2}$ and the mass
of vector mesons, one can study the transition
between the perturbative (hard) 
and nonperturbative (soft) regimes.
One of the interesting 
consequence of the color dipole gBFKL dynamics
is a steeper energy dependence 
of the dipole cross section
at smaller dipole size \cite{NZZ94,NZBFKL}
which can be studied also using
the scanning phenomenon.

Diffractive production of the $2S$ radially excited vector
mesons 
\beq
\gamma^{*}p\rightarrow V'p\,,~~~~~V'(2S)=\rho',\,\omega',\,
\phi',\,\Psi',\,\Upsilon'\,...
\label{eq:1.3}
\endeq
is particularly interesting because of the node effect: 
a strong cancellation
of dipole size contributions to the production amplitude
coming
from the region above and below the node position,
$r_{n}$, in the $2S$ radial wave function \cite{KZ91,NNN92,NNZanom}.
One of the important consequence of the
node effect is a strong
suppression of the photoproduction of radially
excited vector mesons $V'(2S)$
vs. $V(1S)$ mesons
\cite{KZ91,NNN92}. The NMC experiment
\cite{NMCPsi'} and recently the
E687 experiment \cite{E687Psi'} confirmed this suppression
for the 
ratio $\Psi'/(J/\Psi)$.

There are two main reasons which affect the cancellation pattern
in the $V'(2S)$ production amplitude.
The first is connected with the $Q^{2}$ behaviour
of the scanning radius $r_{S}$ (\ref{eq:1.2});
for the electroproduction of $V'(2S)$ light vector mesons
at moderate $Q^{2}$ when the scanning radius
$r_{S}$ is comparable to $r_{n}$,
even a slight variation of
$r_{S}$ with $Q^{2}$ strongly changes the cancellation
pattern  and leads to an anomalous
$Q^{2}$ dependence \cite{KZ91,NNN92,NNZanom}. 
The second reason is due to the
different dipole-size 
dependence of the color dipole cross section 
at different energies in accordance with the gBFKL
dynamics leading also to an 
anomalous energy dependence for the $V'(2S)$
vector meson production. 

In the photoproduction limit of
very small $Q^{2}$, there are two possibilities occurring in the
$V'(2S)$ production amplitude. 
The relative sign of the $V'(2S)$ and
$V(1S)$ production amplitudes can be 
opposite (the overcompensation scenario of ref. \cite{NNZanom}).
The second case corresponds to the same sign of both amplitudes
(the undercompensation scenario of ref. \cite{NNZanom}).
The relative sign of the $V'$ and $V$ production amplitudes
is experimentally measurable using the so-called S\"oding
effect \cite{Soding,Pumplin,Abe}.
 
 
 
\section{Color dipole factorization}

Here we present a short review of the color dipole phenomenology
of diffractive electroproduction of vector mesons developed
in \cite{NNPZ96}.
A meson as a color dipole is described by
the distribution
of the transverse separation, ${\bf{r}}$, of the quark and
antiquark given by the $q\bar{q}$ wave function,
$\Psi({\bf{r}},z)$, where $z$ is
the fraction of the meson lightcone momentum
carried by a quark. 
The interaction of the relativistic
color dipole moment, ${\bf{r}}$, with the
target nucleon is quantified by the energy dependent color
dipole cross section, $\sigma(\nu,r)$.
The Fock state expansion for the 
relativistic meson starts
with the $q\bar{q}$ state and
the higher Fock states $q\bar{q}g...$ 
become very important at high energy. 
In the leading-log ${1\over x}$ approximation, the
effect of higher Fock states can be 
reabsorbed into the energy dependence 
of $\sigma(\nu,r)$, which satisfies 
the generalized BFKL equation 
\cite{NZ94,NZZ94}. 
The dipole cross section is flavour 
independent and represents the universal 
function of $r$ which describes
various diffractive processes in unified form. 
Within this color dipole formalism,
the imaginary part of the production amplitude for the 
virtual photoproduction of vector mesons 
in the forward direction ($t=0$) reads
\beq
{\rm Im}{\cal M}=\langle V|\sigma(\nu,r)|\gamma^{*}\rangle
=\int_{0}^{1} dz \int d^{2}{\bf{r}}\,\sigma(\nu,r)
\Psi_V^{*}({\bf{r}},z)
\Psi_{\gamma^{*}}({\bf{r}},z) \,
\label{eq:2.2}
\endeq
whose normalization is 
$
\left.{d\sigma/ dt}\right|_{t=0}={|{\cal M}|^{2}/ 16\pi}.
$
$\Psi_{\gamma^{*}}(\vec{r},z)$ and
$\Psi_{V}(\vec{r},z)$ represent the
probability amplitudes
to find the color dipole of size $r$
in the photon and quarkonium (vector meson), respectively.
The color dipole distribution in (virtual) photons was derived in
\cite{NZ91,NZ94}.

Eq. (\ref{eq:2.2}) represents the 
color dipole factorization formula because of  
the diagonalization of the scattering
matrix in the $({\bf{r}},z)$ representation.

The energy dependence of the dipole cross section is quantified
by the dimensionless
rapidity, $\xi=\log{1\over x_{eff}}$, where
$x_{eff}= (Q^{2}+m_{V}^{2})/2m_{p}\nu$
and  $m_{V}$ is a mass of the vector meson. 
At large energies corresponding to the HERA energy region, 
the Regge parameter is large,
$\omega=1/x_{eff} \gg 1$, and
pomeron exchange dominates.
 
The more explicit form of the forward production amplitudes for the
transversely (T) and the longitudinally (L) polarized vector mesons
reads \cite{NNZscan}
\arr
{\rm Im}{\cal M}_{T}(x_{eff},Q^{2})=
{N_{c}C_{V}\sqrt{4\pi\alpha_{em}} \over (2\pi)^{2}}
\cdot~~~~~~~~~~~~~~~~~~~~~~~~~~~~~~~~~
\nonumber \\
\cdot \int d^{2}{\bf{r}} \sigma(x_{eff},r)
\int_{0}^{1}{dz \over z(1-z)}\left\{
m_{q}^{2}
K_{0}(\varepsilon r)
\phi(r,z)-
[z^{2}+(1-z)^{2}]\varepsilon K_{1}(\varepsilon r)\partial_{r}
\phi(r,z)\right\}\nonumber \\
 =
{1 \over (m_{V}^{2}+Q^{2})^{2}}
\int {dr^{2} \over r^{2}} {\sigma(x_{eff},r) \over r^{2}}
W_{T}(Q^{2},r^{2})
\label{eq:2.4}
\endarr
\arr
{\rm Im}{\cal M}_{L}(x_{eff},Q^{2})=
{N_{c}C_{V}\sqrt{4\pi\alpha_{em}} \over (2\pi)^{2}}
{2\sqrt{Q^{2}} \over m_{V}}
\cdot~~~~~~~~~~~~~~~~~~~~~~~~~~~~~~~~~
 \nonumber \\
\cdot \int d^{2}{\bf{r}} \sigma(x_{eff},r)
\int_{0}^{1}dz \left\{
[m_{q}^{2}+z(1-z)m_{V}^{2}]
K_{0}(\varepsilon r)
\phi(r,z)-
\varepsilon K_{1}(\varepsilon r)\partial_{r}
\phi(r,z)\right\}\nonumber \\
 =
{1 \over (m_{V}^{2}+Q^{2})^{2}}
{2\sqrt{Q^{2}} \over m_{V}}
\int {dr^{2} \over r^{2}} {\sigma(x_{eff},r) \over r^{2}}
W_{L}(Q^{2},r^{2})
\label{eq:2.5}
\endarr
where
\beq
\varepsilon^{2} = m_{q}^{2}+z(1-z)Q^{2}\,,
\label{eq:2.6}
\endeq
$\alpha_{em}$ is the fine structure 
constant, $N_{c}=3$ is the number of colors,
$C_{V}={1\over \sqrt{2}},\,{1\over 3\sqrt{2}},\,{1\over 3},\,
{2\over 3},\,{1\over 3}~~$ for 
$\rho^{0},\,\omega^{0},\,\phi^{0},\, J/\Psi,\,\Upsilon$ production,
respectively, and
$K_{0,1}(x)$ are the modified Bessel functions.
The detailed discussion and parameterization 
of the lightcone radial wave function, $\phi(r,z)$,
of the $q\bar{q}$ Fock state of the vector meson
is given in \cite{NNPZ96}.
The terms 
$\propto \partial_{r}\phi(r,z)$ in
(\ref{eq:2.4}) and (\ref{eq:2.5}) represent
the relativistic corrections 
which become important
at large $Q^{2}$ and for the production
of light vector mesons.

The real part of the production amplitudes can be included in
Eqs. (\ref{eq:2.4}),(\ref{eq:2.5}) 
using the following substitution \cite{GribMig}
\beq
\sigma(x_{eff},r) \Longrightarrow \left(1-i
\cdot\frac{\pi}{2}\cdot\frac{\partial}
{\partial\,\log\,x_{eff}} \right)\sigma(x_{eff},r)\,. 
\label{eq:2.7}
\endeq 
 
For small $r$,
in the leading-log ${1\over x}$
the dipole cross section can be related to the
gluon structure function, $G(x,q^2)$, of the target nucleon
through
\beq
\sigma(x,r) =
\frac{\pi^2}{3}r^2\alpha_s(r)G(x,q^2) \, ,
\label{eq:2.8}
\endeq
where the gluon structure function enters at
the scale
$q^2 \sim {B\over r^2}$ \cite{Barone} with
$B\sim 10$ \cite{NZglue}.

The integrands of (\ref{eq:2.4}),(\ref{eq:2.5}) are smooth at
small $r$ and decrease exponentially at $r > 1/\epsilon$ due to 
the modified Bessel functions.
Due to the $\sigma(x,r) \propto r^{2}$ behaviour 
(\ref{eq:2.8}), the
amplitudes (\ref{eq:2.4}),(\ref{eq:2.5}) are dominated 
by the dipole size, $r \approx r_{S}$
(scanning phenomenon). 
Then, a
simple evaluation gives \cite{KNNZ94}
\beq
{\rm Im}
{\cal M}_{T} \propto r_{S}^{2}\sigma(x_{eff},r_{S}) \propto
{1 \over Q^2+m_{V}^{2}}\sigma(x_{eff},r_{S}) \propto
 {1\over (Q^{2}+m_{V}^{2})^{2} }
\label{eq:2.9}
\endeq
and
\beq
{\rm Im}
{\cal M}_{L} \approx {\sqrt{Q^{2}}\over m_{V}}{\cal M}_{T}
 \propto
{\sqrt{Q^{2}}\over m_{V}}
 r_{S}^{2}\sigma(x_{eff},r_{S})
 \propto
{\sqrt{Q^{2}}\over m_{V}}
 {1\over (Q^{2}+m_{V}^{2})^{2}}
\label{eq:2.10}
\endeq
respectively. 
All the experiments on $\rho^{0}$ electroproduction
confirm the dominance of the $L$
cross section at large $Q^{2}$ \cite{E665rho,NMCfirho,ZEUSrhoQ2}.
Note, that Eq. (\ref{eq:2.9}) differs from
the familiar vector dominance model (VDM) 
predictions, $M_T \propto {1\over (m_V^2+Q^2)}\sigma_{tot}(\rho N)$.

The scanning phenomenon can be analysed
in terms of the weight functions
\cite{NNZscan},
$W_{T,L}(Q^{2},r^{2})$, 
which are
sharply peaked at $r\approx A_{T,L}/\sqrt{Q^{2}+m_{V}^{2}}$.
At small $Q^{2}$, the values of the scale parameter, $A_{T,L}$, are 
close to $A\sim 6$, which follows from $r_{S}=3/\varepsilon$ with
the nonrelativistic choice $z=0.5$. In general, $A_{T,L} \geq 6$
and increases slowly with $Q^2$ \cite{NNZscan}.
 
For $Q^{2}+m_{V}^{2}\lsim
10-20$\,GeV$^{2}$, the production amplitudes
receive a substantial
contribution from semiperturbative and nonperturbative 
dipole sizes.
In \cite{NZHera,NNZscan} this contribution was modeled by
the energy independent soft cross section, $\sigma^{(npt)}(r)$.
The particular form of this cross section successfully
predicted \cite{NZHera} the proton structure function at very
small $Q^{2}$ recently measured by the E665 collaboration
\cite{E665lowQ2}.
The detailed description
of the dipole cross section used in the present analysis is
given in \cite{NZHera,NNZscan}.
 
 
\section{Diffractive $\rho^{0}$ and $\phi^{0}$
         electroproduction}

The color dipole dynamics predicts
a rapid decrease of production amplitudes
(\ref{eq:2.9}),(\ref{eq:2.10}) at large $Q^{2}$.
Fig.~1 presents our predictions for $\rho^{0}$ and $\phi^{0}$
electroproduction together 
with the NMC data \cite{NMCfirho} and 
the data from the HERA experiments \cite{ZEUSrhoQ2,H1rhoQ2}. 
Here, the $Q^{2}$ dependence of 
the observed polarization-unseparated total production
cross section, $\sigma(\gamma^{*}\rightarrow V)=
\sigma_{T}(\gamma^{*}\rightarrow V)+
\epsilon \sigma_{L}(\gamma^{*}\rightarrow V)$, 
is shown for the value of the $L$ polarization
of the virtual photon,
$\epsilon$,  taken from the corresponding
experiment.

In adition to the pure pomeron exchange contribution to the production
amplitude, the secondary Reggeon exchanges can also be 
important but not at HERA energies, where
the Regge parameter, $\omega$, is very large and
Eqs. (\ref{eq:2.4}),(\ref{eq:2.5}) can be used for
a description of electroproduction of vector mesons at
high energy.
Not so at the lower energy of the NMC experiment.
The fit to $\sigma_{tot}(\gamma p)$ can, for instance, 
be cast in the form
$\sigma_{tot}(\gamma p) = \sigma_{\Pom}(\gamma p)\cdot
\left(1 + A/\omega^{\Delta}\right)$,
where the term $A/\omega^{\Delta}$ in the factor
$f = 1 + A/\omega^{\Delta}$ represents the non-vacuum
Reggeon exchange contribution.
The Donnachie-Landshoff fit gives
$A = 2.332$ and $\Delta = 0.533$ \cite{DLRegge}.
The application of this
correction,  
$\sigma(\gamma^{*} \rightarrow \rho^{0}) =
f^{2} \sigma_{\Pom}(\gamma^{*} \rightarrow \rho^{0})$,
brings the theory to a better agreement with the NMC data.
For $\phi^{0}$ production, $f\equiv 1$ due to the Zweig rule and
the pure pomeron contribution correctly describes
the NMC data \cite{NMCfirho}.

Another important prediction of the
gBFKL dynamics is
a steeper rise with energy of the production cross section,
$\sigma(\gamma^{*}\rightarrow V)$, at higher $Q^{2}$
and/or for heavy quarkonia \cite{NNZscan}.
The high-energy predictions of the model 
for the production cross section
are in good agreement with the HERA data 
for $Q^{2}=0$ (Fig.~2) as well as for large $Q^{2}$ 
(Fig.~1). This confirms the growth
of the dipole cross section with energy expected from the gBFKL
dynamics.

Fig.~2 represents our predictions for the energy dependence
of real $\phi^{0}$ and $\rho^{0}$
photoproduction 
(with and without secondary Reggeon corrections for $\rho^{0}$ 
production).
The Reggeon correction factor, $f^{2}$,
brings the theory to a better agreement with the low energy
$\rho^{0}$ production data \cite{Rholownu}.
Our predictions for high energy
agree well with the recent 
ZEUS data \cite{ZEUSrho94,ZEUSrho95}.
We find also good agreement with the fixed target
\cite{Philownu}
and ZEUS \cite{ZEUSphi}
data on real $\phi^{0}$ photoproduction.

A smaller scanning radius
for $\phi^{0}$ photoproduction vs.
$\rho^{0}$ photoproduction results in
a steeper energy
dependence of the former (see Fig.2).
At $W=70$\,GeV, we predict
$\sigma(\gamma\rightarrow \phi^{0})=0.87 \mu b$
which agrees with the first ZEUS measurement
$\sigma(\gamma\rightarrow\phi^{0})=0.95 \pm 0.33$\,$\mu b$ 
\cite{ZEUSphi}.

Fig.3 represents our predictions for
$R_{LT} = d\sigma_{L}(\gamma^{*} \rightarrow
V)/d\sigma_{T}(\gamma^{*} \rightarrow V)$ $\cdot m_{V}^{2}/Q^{2}$.
The steady decrease of $R_{LT}$ with $Q^{2}$ 
reflects
a larger contribution from large-size dipoles to the $T$
production amplitude,
i.e., $A_{T} \gsim A_{L}$ \cite{NNZscan}
as a very
specific prediction of the color dipole
approach. 
The available experimental data
\cite{E665rho,NMCfirho,ZEUSrhoQ2} 
confirm $R_{LT} < 1$ but
the error bars ate still quite large.


 
\section{Anomalies in the electroproduction of $2S$ radially
         excited vector mesons}

Here the keyword is the node effect - the
$Q^{2}$ and energy dependent cancellations
from the large and small size contributions
to the production amplitude of the $V'(2S)$ vector
mesons.
The $Q^{2}$ and energy dependence of the node
effect follows from the $Q^{2}$ dependence of the
scanning radius (which is close to the
node position $r_{n}\sim R_{V}$) and
from the
different energy dependence of the dipole cross section at
small ($r<R_{V}$) and large ($r>R_{V}$) dipole sizes.
In this case the predictive power becomes very weak.

In the nonrelativistic limit of heavy quarkonia, the
node effect does not depend on the polarization of the virtual photon
and of the produced vector meson. Not so for light
vector mesons due to the
different wave functions for
the $T$ and $L$
polarized photons and to the fact that
different regions of $z$ 
contribute to the ${\cal M}_{T}$ and ${\cal M}_{L}$. 

Two cases can occur in the $2S$ production amplitude; the undercompensation
and the overcompensation scenario \cite{NNZanom}.
In the undercompensation case,
the $2S$ production amplitude
is dominated by the positive contribution coming from small
dipole sizes $r\lsim r_{n}$
and the $V(1S)$ and $V'(2S)$ photoproduction
amplitudes have the same sign. 
With our model wave functions this scenario is realized for 
$T$ polarized $\rho'(2S)$ and $\phi'(2S)$.
In this scenario, a decrease of the scanning radius 
with $Q^{2}$ leads to a rapid
rise of the $V'(2S)/V(1S)$ production ratio with $Q^{2}$
\cite{NNZanom}, see Fig.~4; then at
$Q^{2}\gsim 1$\,GeV$^{2}$
the $V'(2S)$ and $V(1S)$
production cross sections become comparable,
when the production
amplitudes are dominated by 
a dipole size $r\ll r_{n}$ \cite{NNZanom,NNZscan}.

An interesting situation occurs in the production of
$L$ polarized $\rho'(2S)$ and $\phi'(2S)$ mesons, where
our model wave functions predict overcompensation;
at $Q^{2}=0$\,GeV$^{2}$ the amplitude is dominated by
the negative contribution
coming from large dipole sizes $r\gsim r_{n}$.
Consequently,
with the increase of $Q^{2}$, the scanning radius
decreases and one has the {\sl exact} cancellation
of the large and small dipole size contributions
to the production amplitude. 
We find this exact node effect
at some value
$Q_{n}^{2}\sim 0.5$\,GeV$^{2}$ for both  $\rho'(2S)$ and
$\phi'(2S)$ production (see Fig.~3). 

Decreasing further
$r_{S}$, the overcompensation scenarion goes into 
the above described
undercompensation one and
for both the $T$ and $L$
polarized mesons we
predict a steep rise with $Q^{2}$ of the
$V'(2S)/V(1S)$ ratios
on the scale $Q^{2}\sim 0.5$\,GeV$^{2}$. 
At large $Q^{2}$ where the production of
$L$ polarized mesons dominates,  the
$\rho'(2S)/\rho^{0}(1S)$ and $\phi'(2S)/\phi^{0}(1S)$ cross section ratios
level off at $\sim 0.3$ (see Fig.~4).
Due to the
different node effect for the $T$ and $L$ polarizations,
we find $R_{LT}(2S)\ll R_{LT}(1S)$ , see Fig.~3.
 
Fig.~5 represents the color dipole model prediction   
for the $Q^{2}$ dependence of the
polarization-unseparated forward cross section ratios
$\sigma(\gamma^{*}\rightarrow \rho'(2S))/
\sigma(\gamma^{*}\rightarrow \rho^{0})$
and
$\sigma(\gamma^{*}\rightarrow \phi'(2S))
/\sigma(\gamma^{*}\rightarrow \phi^{0})$ at $W=100$\,GeV.
The
anomalous properties of $\sigma_{L}$ (due to its smallness)
at small $Q^{2}$
are essentially
invisible in the polarization-unseparated $V'(2S)$ production
cross section shown in Fig.~5.

For the $L$ polarized $V'(2S)$ we have 
an onset of the overcompensation scenario. 
At moderate energy and
$Q^{2}$ very close to but
smaller then $Q_{n}^{2}$, the negative 
contribution from $r\gsim r_{n}$ takes over in the $V'(2S)$
production amplitude. 
Because of a steeper energy rise of the dipole cross
section at smaller dipole sizes, the positive
contribution to the production amplitude coming from the region
below the node position rises faster with energy
and gradually takes over.
At some 
intermediate energy, we find an exact cancellation
of these two contributions to the production
amplitude and a minimum of the $V'(2S)$ production cross section. 
Fig.~4 shows such a nonmonotonic energy dependence
of the $\rho'(2S)$ and $\phi'(2S)$ production at 
$Q^{2}\approx 0.5$\,GeV$^{2}$ which corresponds to our model
wave functions.
At higher $Q^{2}$ and smaller
scanning radii $r_{S}$, we predict very weak 
energy dependence of the $V_{L}(2S)/V_{L}(1S)$
production ratio.
 
 
\section{ Conclusions}

We have
presented the phenomenology of diffractive
photo- and electroproduction of $1S$ and
$2S$ vector mesons in the framework of the color dipole
gBFKL dynamics.
There are two main aspects 
of vector meson production
coming from the gBFKL dynamics.
First, the energy dependence of the $1S$ vector meson production
is controlled by the energy dependence of the dipole cross
section which is steeper for smaller dipole sizes.
This results in a steeper rise with energy of the production
cross section at higher $Q^{2}$ and/or for heavy vector mesons.
Second,
the 
$Q^{2}$ dependence of the $1S$ vector meson production is
controlled by the shrinkage of the transverse size of the virtual
photon and the small-size dependence of the color dipole cross
section.
We present a good quantitative description of
the experimental data on diffractive $\rho^{0}$ and $\phi^{0}$ 
photo- and electroproduction which confirms
the consequences of the gBFKL dynamics mentioned above.

For the production of
the $V'(2S)$ radially excited vector mesons, we predict 
a rich pattern of anomalous $Q^{2}$ and energy dependence
as compared to a smooth $Q^{2}$ and energy dependence for
the $V(1S)$ ground state vector mesons.
These anomalies come from the node
in the $2S$ radial wave function in conjuction with
the scanning phenomenon and
the energy dependence of the dipole cross section. 
We predict a very strong suppression of the
$V'(2S)/V(1S)$ production ratio in the real photoproduction limit.
For the production of $L$ polarized $2S$
mesons we find an overcompensation scenario leading to a sharp 
dip in the production cross section at some finite $Q^{2}
=Q_{n}^{2}\sim 0.5$\,GeV$^{2}$. 
The position of this dip
is energy dependent and leads to a nonmonotonic energy
dependence of $\sigma_{L}(2S)$ at fixed $Q^{2}$.
The relative sign of the $\rho'$ and $\rho^{0}$ 
production amplitude
can be measured directly using the
S\"oding-Pumplin method.
At larger $Q^{2}$, i.e. at smaller $r_{S}$,
the $2S/1S$ cross section ratio rises steeply on the
scale $Q^{2} \lsim 0.5\,GeV^{2}$.
At large $Q^{2}$, we find the flattening of this $2S/1S$ ratio as
a non-negotiable prediction from the color dipole dynamics.

\pagebreak
{\bf Figure captions:}
\begin{itemize}

\item[Fig.~1]
~- The color dipole model
predictions for the $Q^2$ dependence of the observed
cross section $\sigma(\gamma^{*}\rightarrow V)=
\sigma_{T}(\gamma^{*}\rightarrow V)+\epsilon
\sigma_{L}(\gamma^{*}\rightarrow V)$
of exclusive $\rho^0$ and $\phi^0$
production 
compared with the low-energy NMC \cite{NMCfirho}
and high-energy  ZEUS
\cite{ZEUSrhoQ2} and H1 \cite{H1rhoQ2} data.
The top curve is a prediction for the $\rho^{0}$
production at $W=70$\,GeV, the lower curves are for the
$\rho^{0},\phi^{0}$ production at $W=15$\,GeV.
The dashed curve (for $\rho^{0}$) shows the pure pomeron contribution
$\sigma_{\Pom}(\gamma^{*}\rightarrow \rho^{0})$, while the
solid curve (for $\rho^{0}$)
shows the effect of correcting for the non-vacuum
Reggeon exchange as described in the text.

\item[Fig.~2]
~- The color dipole model
predictions for the  energy dependence of real photoproduction
of the $\phi^{0}$ mesons compared with fixed target
\cite{Philownu} and high energy  ZEUS  data (open square
for the $\phi^{0}$ \cite{ZEUSphi}, solid circle for the
$\rho^{0}$ \cite{ZEUSrho94,ZEUSrho95}).

\item[Fig.~3]
~- The color dipole model
predictions for the $Q^2$ and $\nu$ dependence of the
ratio of the longitudinal and transverse differential cross
sections in the form of the quantity
$
R_{LT}={m_{V}^{2} \over Q^{2}}{d\sigma_{L}(\gamma^{*}\rightarrow V)
\over d\sigma_{T}(\gamma^{*}\rightarrow V)}\,,
$
where $m_{V}$ is the mass of the vector meson.
The solid and dashed curves are for $W=15\,GeV$ and $W=150\,GeV$.

\item[Fig.~4]
~- The color dipole model
predictions for the $Q^2$ and $W$ dependence of the ratios
$\sigma(\gamma^{*}\rightarrow \rho'(2S))/
\sigma(\gamma^{*}\rightarrow \rho^{0})$ and
$\sigma(\gamma^{*}\rightarrow \phi'(2S))/
\sigma(\gamma^{*}\rightarrow \phi^{0})$
for the (T) and (L)
polarization of the vector mesons.

\item[Fig.~5]
~- The color dipole model
predictions for the $Q^{2}$ dependence of the ratio of
the polarization-unseparated forward production cross sections
$d\sigma(\gamma^{*}\rightarrow \rho'(2S))/
d\sigma(\gamma^{*}\rightarrow \rho^{0})$ and
$d\sigma(\gamma^{*}\rightarrow \phi'(2S))/
d\sigma(\gamma^{*}\rightarrow \phi^{0})$
for the polarization of
the virtual photon $\epsilon = 1$ at the HERA energy $W=100
\,GeV$.

\end{itemize}

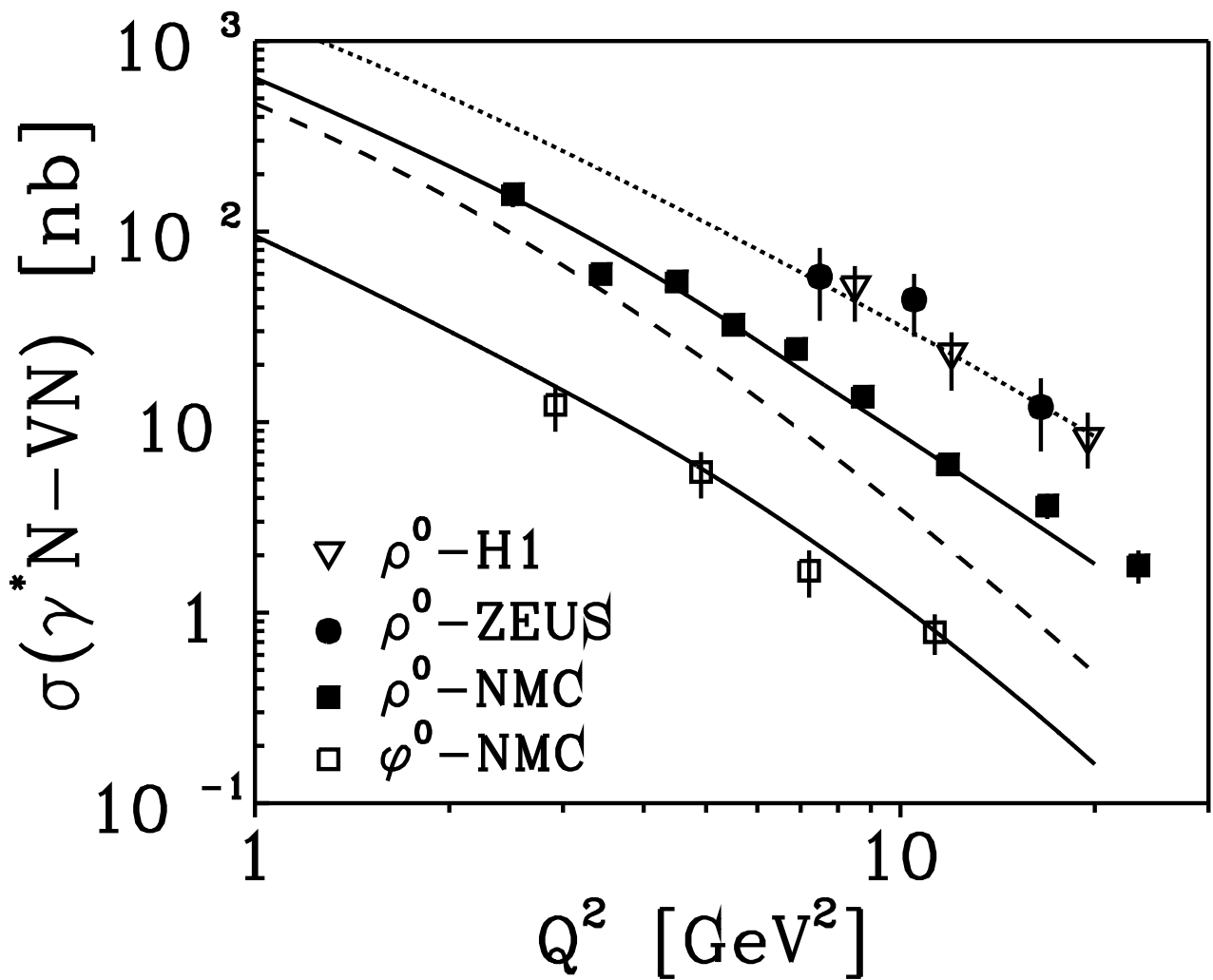
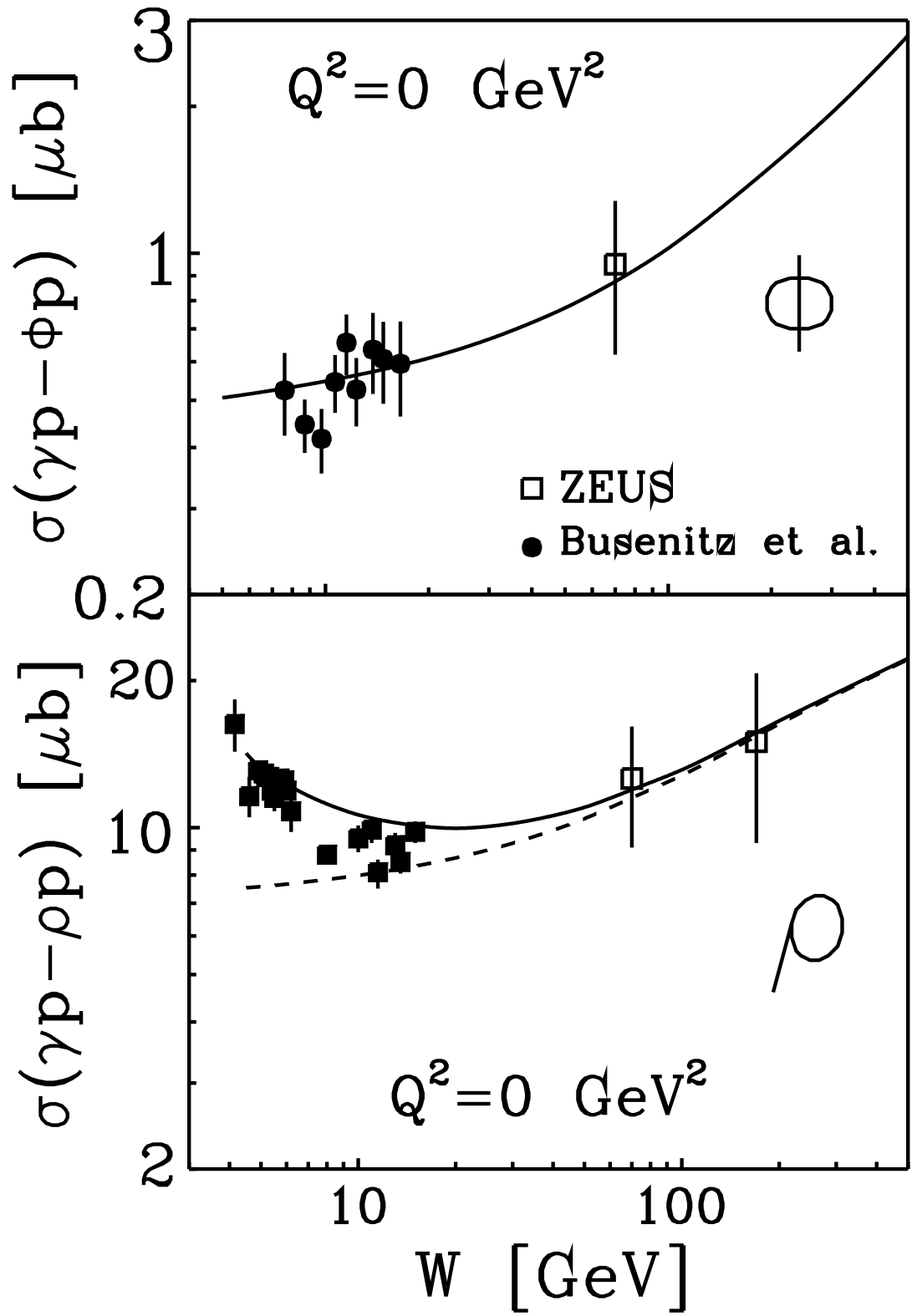
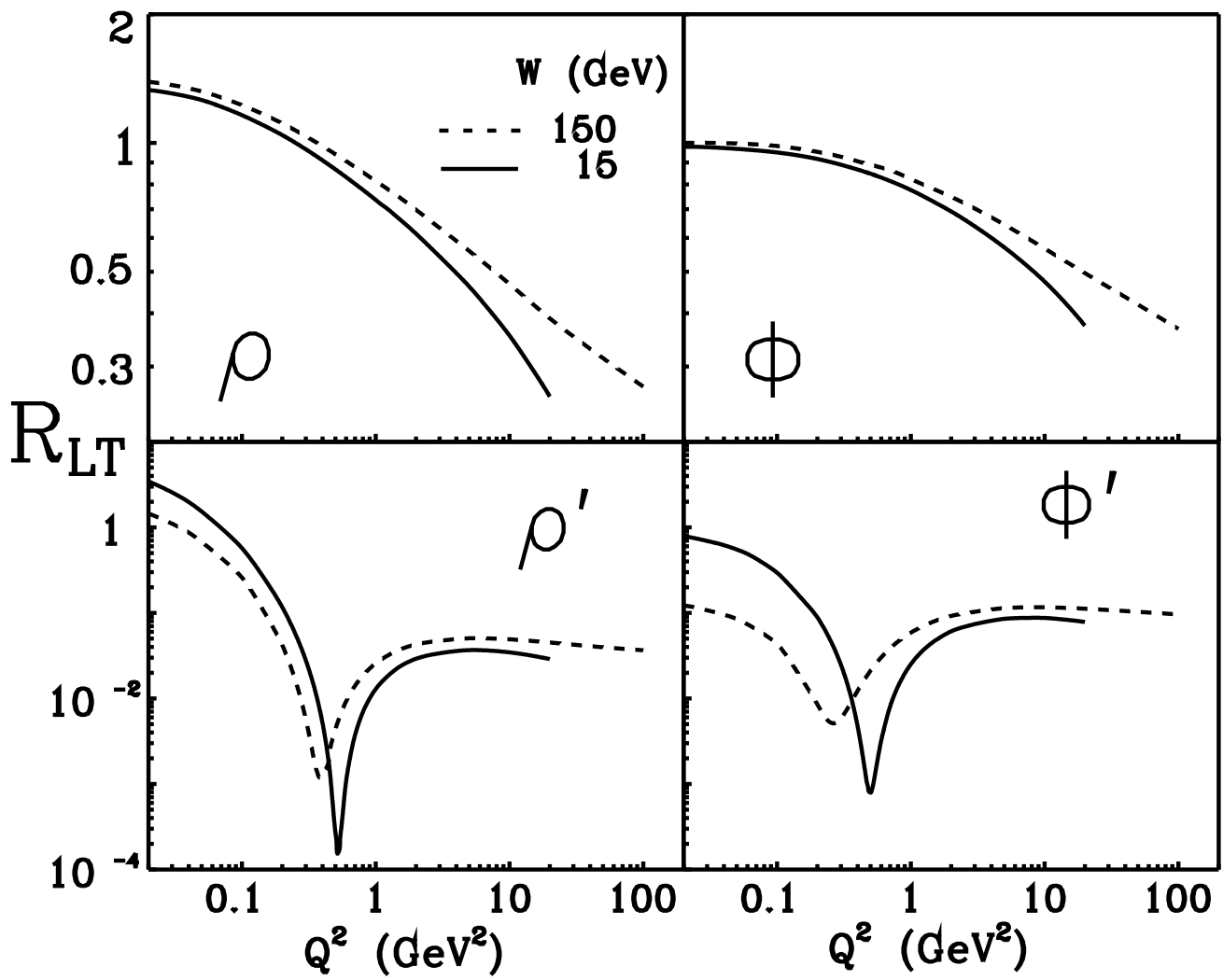
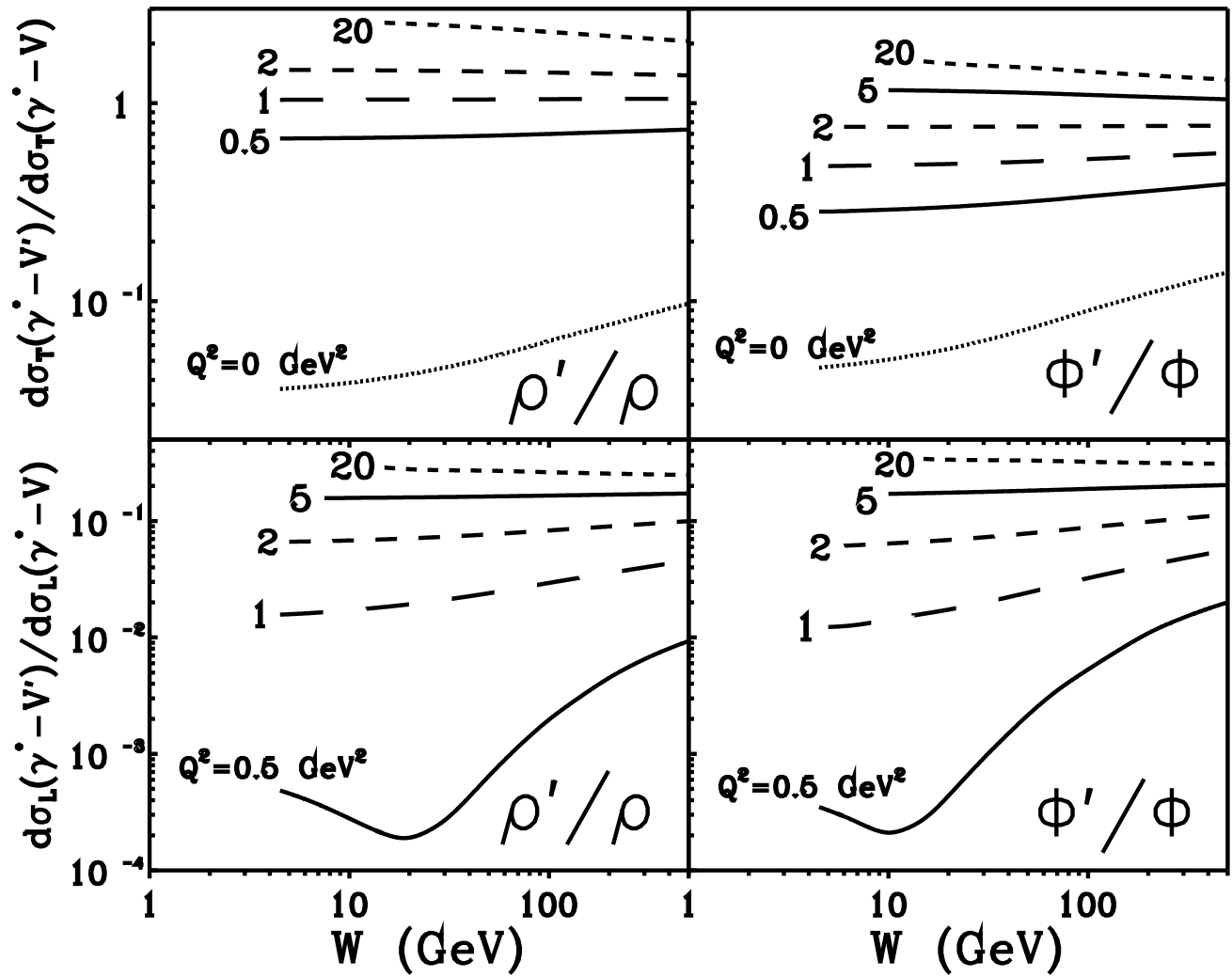
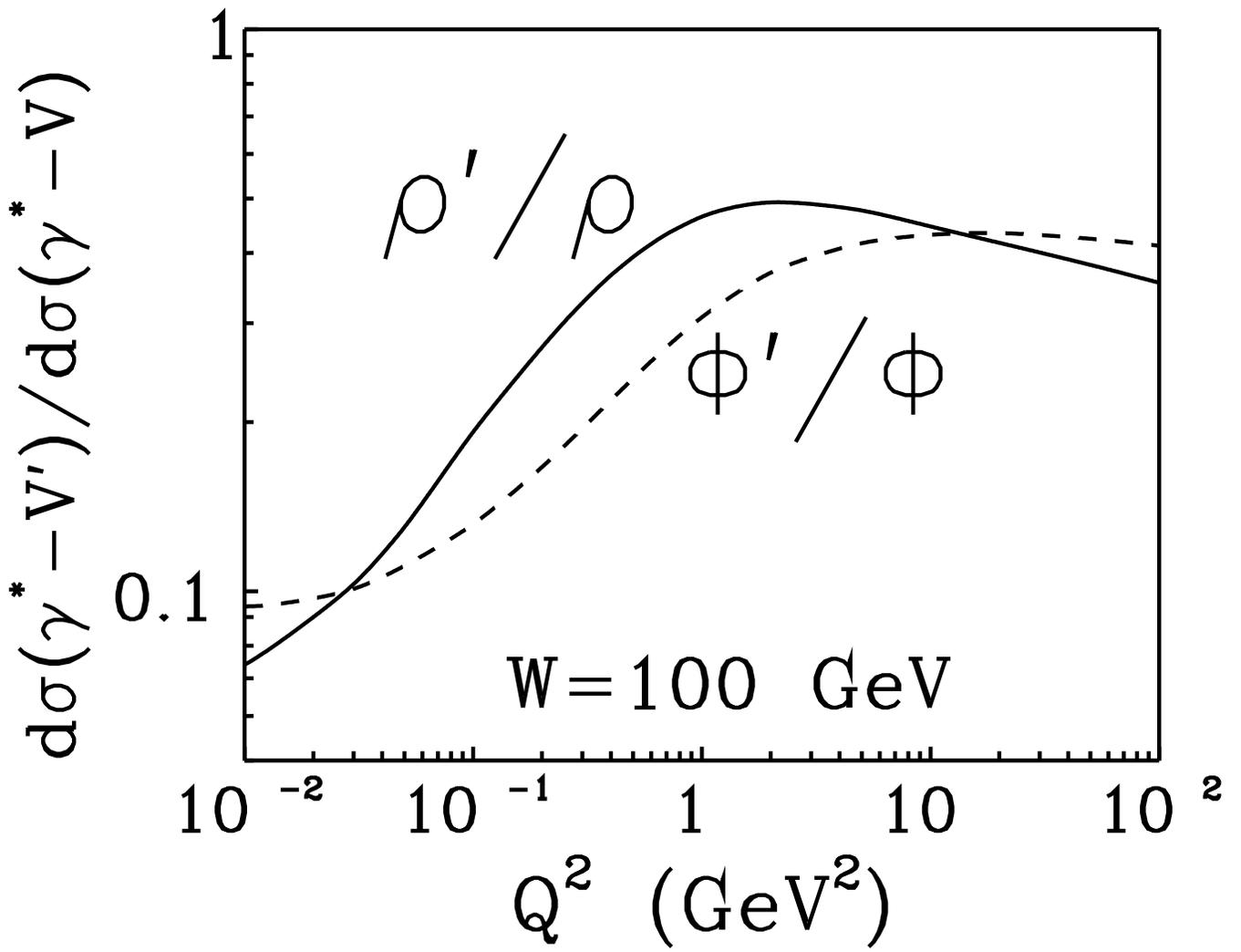
\end{document}